# Physics-Informed Deep Learning for Solving Phonon Boltzmann Transport Equation with Large Temperature Non-Equilibrium

Ruiyang Li[1], Jian-Xun Wang[1], Eungkyu Lee[2]*, and Tengfei Luo[1,3,4]*

[1] Department of Aerospace and Mechanical Engineering, University of Notre Dame, Notre Dame, Indiana 46556, USA

[2] Department of Electronic Engineering, Kyung Hee University, Yongin-si, Gyeonggi-do 17104, South Korea

[3] Department of Chemical and Biomolecular Engineering, University of Notre Dame, Notre Dame, Indiana 46556, USA

[4] Center for Sustainable Energy of Notre Dame (ND Energy), University of Notre Dame, Notre Dame, Indiana 46556, USA

* Corresponding author. Email: eleest@khu.ac.kr (E. Lee); tluo@nd.edu (T. Luo)

**Abstract**

Phonon Boltzmann transport equation (BTE) is a key tool for modeling multiscale phonon transport, which is critical to the thermal management of miniaturized integrated circuits, but assumptions about the system temperatures (*i.e.*, small temperature gradients) are usually made to ensure that it is computationally tractable. To include the effects of large temperature non-equilibrium, we demonstrate a data-free deep learning scheme, physics-informed neural network (PINN), for solving stationary, mode-resolved phonon BTE with arbitrary temperature gradients. This scheme uses the temperature-dependent phonon relaxation times and learns the solutions in parameterized spaces with both length scale and temperature gradient treated as input variables. Numerical experiments suggest that the proposed PINN can accurately predict phonon transport (from 1D to 3D) under arbitrary temperature gradients. Moreover, the proposed scheme shows great promises in simulating device-level phonon heat conduction efficiently and can be potentially used for thermal design.

## Introduction

Multiscale phonon transport from the ballistic limit to the diffusive extreme is ubiquitous in technologically important materials and applications, such as thermoelectrics[1,2] and miniaturized electronic systems like central processing unit and flat-panel display[3,4]. A better understanding of thermal transport mechanisms in a large span of length scales is critical to engineering such materials and devices to achieve better properties[1] and performances[5]. Phonon Boltzmann transport equation (BTE)[6,7] is a method capable of modeling phonon transport from ballistic to diffusive regimes, and a large number of research efforts have been made to devise numerical solvers for it[8]. However, due to the limited computational efficacy of existing methods, phonon BTE faces difficulties when applied to complicated problems involving mode-resolved phonon properties, high spatial dimensions, and large temperature non-equilibrium. It is thus desirable to develop an accurate and efficient mode-resolved BTE solver for predicting heat conduction particularly under various temperature ranges and length scales, where phonons can follow very different transport mechanisms[9,10].

Due to the challenges related to computational cost, apart from the commonly adopted assumptions like single-mode relaxation time approximation and isotropic phonon dispersion[6], the temperature difference is usually assumed to be sufficiently small (relative to a reference temperature) to simplify the computations[11-14]. With small temperature differences, the phonon equilibrium deviational distribution can be linearly approximated by temperature[15,16], and the relaxation time is often treated as temperature independent and thus spatially invariant. Although success has been achieved in investigating nanoscale size effects for phonon transport under these assumptions[17-19], the results and conclusions cannot be simply generalized to the cases with large temperature differences, since the phonon relaxation time or mean free path does depend on the temperature[9,10]. Even in the fully diffusive limit, it is well known that as the local temperature changes drastically, the thermal conductivity can vary across the system domain. Moreover, under large temperature gradients, phonon transport can be in different regimes (diffusive or ballistic) for a given phonon frequency and polarization at different spatial locations[20]. In many applications, hotspots with temperatures much higher than the average system temperature can emerge, such as those in laser material processing[21] and power electronics[22,23], especially at cryogenic temperatures[24]. As a result, it is necessary to develop the capability to model the effects of large temperature differences in phonon BTE.

Despite the necessity of developing a robust solver for phonon BTE under arbitrary temperature differences, it is a challenging task as the BTE can be multiscale in both the frequency and spatial domains. The temperature-dependent relaxation time further adds to the difficulty of solving phonon BTE given the high dimensionality of this partial differential equation (PDE). Numerical methods have been proposed for this problem, such as the Monte Carlo (MC) method[25-27] and deterministic discretization-based methods[28-30]. However, traditional MC methods suffer from statistical errors and become inefficient at small Knudsen numbers (Kn) due to its restrictions on time step and grid size[26,31]. While variance reduction techniques have been employed to enable fast MC simulations, they are only suitable for problems with small deviations from the equilibrium, and thus the computational speedup can only be achieved under near-equilibrium conditions (i.e., small temperature differences)[11,32]. As a widely used deterministic solver, discrete ordinate method (DOM) discretizes the angular space into small solid angles to capture the non-equilibrium phonon distribution. However, DOM and its variants usually converge slowly in the diffusive regime and require large memory[33]. Recently, a finite-volume discrete unified gas kinetic scheme (DUGKS)[34] has been developed for arbitrary temperature difference, but the explicit scheme is known to be restricted by the Courant-Friedrichs-Lewy condition and not efficient for real three-dimensional steady-state problems. In general, few of these methods are both accurate and efficient in predicting phonon heat conduction under arbitrary temperature gradients.

Machine learning-based techniques started to play a role in studying and predicting physical properties in the past decade[35-40]. Deep learning has shown great potentials in solving high-dimensional PDEs to describe the unknown or unrepresented physics[41-45]. Recently, we have developed a deep neural network (DNN) framework for solving stationary mode-resolved phonon BTE[46]. The model can be trained by minimizing the BTE residuals to obey physical laws governed by the BTE without the need of any labeled data. Such a physics-informed neural network (PINN) can return accurate results in the domain of interest very efficiently. Different from other numerical methods, PINNs are trained to approximate a high-dimensional solution function of the phonon BTE by leveraging its capability as a universal function approximator[47]. The evaluation of such trained models can be very fast as the feedforward algorithm only involves a few matrix multiplications. Moreover, parametric learning has also been enabled by treating system parameters (e.g., system size) as additional inputs besides mode-resolved phonon properties, providing a significant advantage of investigating effects of parameters like characteristic length scale, which is a key to determining the ballistic and diffusive nature of the phonon transport process. However, as the first demonstration of PINN for phonon BTE, small temperature difference was still assumed[46]. To make this efficient tool more generally applicable, it is necessary to extend the PINN model so that it can handle problems with large temperature gradients.

In this work, we develop a data-free PINN scheme for solving stationary mode-resolved phonon BTE with arbitrary temperature difference. This scheme uses the temperature-dependent relaxation times and learns the solutions by minimizing the residuals of the governing equations and boundary conditions. Numerical experiments are conducted to validate the model with up to three spatial dimensions. We show that both the length scale and boundary temperature difference can be used as input variables to learn BTE solutions in parameterized spaces, so that a single training can enable the model to be used for evaluating thermal transport at any length scale or temperature difference. We also confirm the effects of large temperature variations, which are difficult to capture in conventional numerical methods. The proposed method performs well in accuracy and efficiency, providing a powerful tool for simulating device-level phonon heat conduction.

## Results

### *Phonon Boltzmann transport equation*

Under the single-mode relaxation time approximation, the mode-resolved phonon BTE at the steady state can be written as[6],

$$\mathbf{v} \cdot \nabla f = \frac{f^{\mathrm{eq}}(T) - f}{\tau(T)}, \tag{1}$$

where $f = f(\mathbf{x}, \mathbf{s}, k, p)$ (or $f(\mathbf{x}, \mathbf{s}, \omega, p)$) is the phonon distribution function dependent on the spatial vector $\mathbf{x}$, directional unit vector $\mathbf{s} = (cos\theta, sin\theta cos\varphi, sin\theta sin\varphi)$ ($\theta$ is the polar angle and $\varphi$ is the azimuthal angle), wave number $k$ (or angular frequency $\omega = \omega(k, p)$) and polarization $p$, and $\mathbf{v}$ is the phonon group velocity. $f^{\mathrm{eq}}$ represents the phonon equilibrium distribution following the Bose-Einstein distribution,

$$f^{\mathrm{eq}}(\omega, p, T) = 1/(e^{\frac{\hbar\omega}{k_\mathrm{B}T}} - 1), \tag{2}$$

where $\hbar$ is the reduced Planck's constant, and $k_\mathrm{B}$ is the Boltzmann constant. It is noted that in Eq. (1) the relaxation time $\tau = \tau(\omega, p, T)$ also depends on the local temperature $T$, meaning that $\tau$ changes spatially across the system for a given phonon frequency and polarization.

In the case of a system without any internal heat source, we have a physical constraint that the divergence of the heat flux **q** must be zero, which can be obtained by integrating the energy-based form of Eq. (1) over the solid angle space ($\Omega$) and the frequency space ($\omega, p$)

$$\nabla \cdot \mathbf{q} = \sum_p \int_0^{\omega_{\max,p}} \int_{4\pi} \hbar\omega D \frac{f^{eq}(T) - f}{\tau(T)} d\Omega\, d\omega = 0, \tag{3}$$

where the heat flux is

$$\mathbf{q} = \sum_p \int_0^{\omega_{\max,p}} \int_{4\pi} \mathbf{v}\hbar\omega D f\, d\Omega\, d\omega, \tag{4}$$

with $D = D(\omega, p)$ and $\omega_{\max,p}$ being the phonon density of states and maximum frequency, respectively.

*PINNs for stationary phonon BTE with arbitrary temperature difference*

For a multiscale thermal transport problem at the steady state, the physical constraints can be expressed as

$$\mathcal{R}(f(\mathbf{x}, \mathbf{s}, k, p, \boldsymbol{\mu}), T) := \begin{cases} \mathbf{v} \cdot \nabla f - \dfrac{f^{eq}(T) - f}{\tau(T)} = 0 \\ \sum_p \int_0^{\omega_{\max,p}} \int_{4\pi} \hbar\omega D \dfrac{f^{eq}(T) - f}{\tau(T)} d\Omega\, d\omega = 0, \end{cases} \quad \mathbf{x}, \mathbf{s}, k, p \in \Gamma, \boldsymbol{\mu} \in \mathbb{R}^d. \tag{5}$$

Here, the phonon distribution *f* is a function of variables in domain $\Gamma$ and additional parameters $\boldsymbol{\mu}$. The solution of *f* can be uniquely determined under certain boundary conditions,

$$\mathcal{B}_i(\mathbf{x}, \mathbf{s}, k, p, f, \boldsymbol{\mu}) = 0, \quad \mathbf{x}, \mathbf{s}, k, p \in \Gamma_b, \boldsymbol{\mu} \in \mathbb{R}^d, \tag{6}$$

where $\mathcal{B}_i$ are the boundary condition operators, and $\Gamma_b$ denotes the boundary region. In the Methods section, we show three typical boundary conditions encountered in phonon BTE, including the isothermal boundary, the diffusely reflecting boundary, and the periodic boundary.

For fast predictions of steady-state multiscale thermal transport with arbitrary temperature difference, a PINN model is developed as depicted in Fig. 1. The input layer is composed of **x**, **s**, $k$, $p$, and parameters of interest **μ**. Parameter **μ** in this work is set to be either length scale $L$ or boundary temperature difference $\Delta T$. We use two fully-connected DNNs to approximate the equilibrium part $f^{\text{eq}}(T)$ and the non-equilibrium part $f^{\text{neq}} = f - f^{\text{eq}}(T)$ of the phonon distribution. Each sub-network maps the inputs to a target output, through several layers of neurons comprising affine linear transformations and scalar nonlinear activation functions. Specifically, the output from one DNN is the equilibrium temperature $T$, which determines $f^{\text{eq}}(T)$ and $\tau(T)$ accordingly. The other output is $f^{\text{neq}}$, and we combine it with $f^{\text{eq}}(T)$ to obtain the total phonon distribution function $f$. While the loss function can be explicitly defined with the residuals of Eqs. (5) and (6), it is difficult to directly minimize the second integral in Eq. (5) as proper nondimensionalization must be performed to evaluate it relative to some appropriate unit. Inspired by the way of linearly approximating $f^{\text{eq}}$ under small temperature difference, an additional shallow neural network (NN) with only one hidden layer is pretrained to generate a scaling factor $\beta(T)$ such that we have

$$f^{\text{eq}}(T) \approx f^{\text{eq}}(T_{\text{ref}}) + \beta(T)(T - T_{\text{ref}}),  \quad (7)$$

where $T_{\text{ref}}$ is the reference temperature. Same as $f^{\text{eq}}(T)$, the scaling factor $\beta(T)$ is also implicitly dependent on ($\omega$, $p$). It is noted that this factor is reduced to $\frac{\partial f^{\text{eq}}}{\partial T}$ under the assumption of small temperature differences. Although Eq. (7) is still a nonlinear approximation of $f^{\text{eq}}$ by $T$, substituting it into Eq. (3) allows us to close the system of equations. If we denote $T$ in the linear part of Eq. (7) as $T^*$ and plug Eq. (7) into Eq. (3), we have

$$T^* = T_{\text{ref}} + \frac{1}{4\pi}\left(\sum_p \int_0^{\omega_{\text{max},p}} \int_{4\pi} \hbar\omega D \frac{f - f^{\text{eq}}(T_{\text{ref}})}{\tau(T)} d\Omega \, d\omega\right) \times \left(\sum_p \int_0^{\omega_{\text{max},p}} \frac{\hbar\omega D \beta(T)}{\tau(T)} d\omega\right)^{-1}. \quad (8)$$

Apparently, $T$ and $T^*$ should be identical. Then Eq. (5) becomes

$$\mathcal{R}(f(\mathbf{x}, \mathbf{s}, k, p, \boldsymbol{\mu}), T) := \begin{cases} \mathbf{v} \cdot \nabla f - \dfrac{f^{\text{eq}}(T) - f}{\tau(T)} = 0 \\ T^* - T = 0, \end{cases} \quad \mathbf{x}, \mathbf{s}, k, p \in \Gamma, \boldsymbol{\mu} \in \mathbb{R}^d, \tag{9}$$

where $T^*$ is calculated based on Eq. (8). The difference between these two temperatures can be easily nondimensionalized by the temperature difference across the simulation domain. Different from conventional solvers that determine the local equilibrium temperature by dichotomy or Newton's method, the introduction of $\beta$ enables efficient computation of $T^*$ without any iterative process. It is also noted that pretraining must be conducted to learn $\beta$ with a shallow NN, which is sufficient to capture the nonlinear relationship between $\beta$ and ($\omega$, $p$, $T$). Therefore, instead of directly using Eq. (2), in the present scheme we compute $f^{\text{eq}}(T)$ based on Eq. (7) with the outputs $\beta$ and $T$ from the two networks (see Fig. 1).

As shown in Fig. 1, given $T_{\text{ref}}$ and the temperature range $T \in [T_{\text{ref}} - \Delta T, T_{\text{ref}} + \Delta T]$, a shallow NN is first trained to provide the scaling factor $\beta = \beta(\omega, p, T)$. Then two DNNs are trained by minimizing the sum of residuals in Eqs. (6) and (9) as follows:

$$\mathcal{L}(\mathbf{W}, \mathbf{b}) = \left\| \mathbf{v} \cdot \nabla f - \frac{f^{\text{eq}} - f}{\tau} \right\|^2 + \|T^* - T\|^2 + \sum_i \|\mathcal{B}_i\|^2, \tag{10}$$

where $\mathbf{W}$ and $\mathbf{b}$ refer to the weights and biases of the entire network, and $\|\cdot\|$ is $L_2$ norm. An optimal set of network parameters can be obtained by minimizing this composite loss function,

$$\mathbf{W}^*, \mathbf{b}^* = \arg\min_{\mathbf{W}, \mathbf{b}} \mathcal{L}(\mathbf{W}, \mathbf{b}). \tag{11}$$

The specific DNN architectures and training details are included in the Methods section.

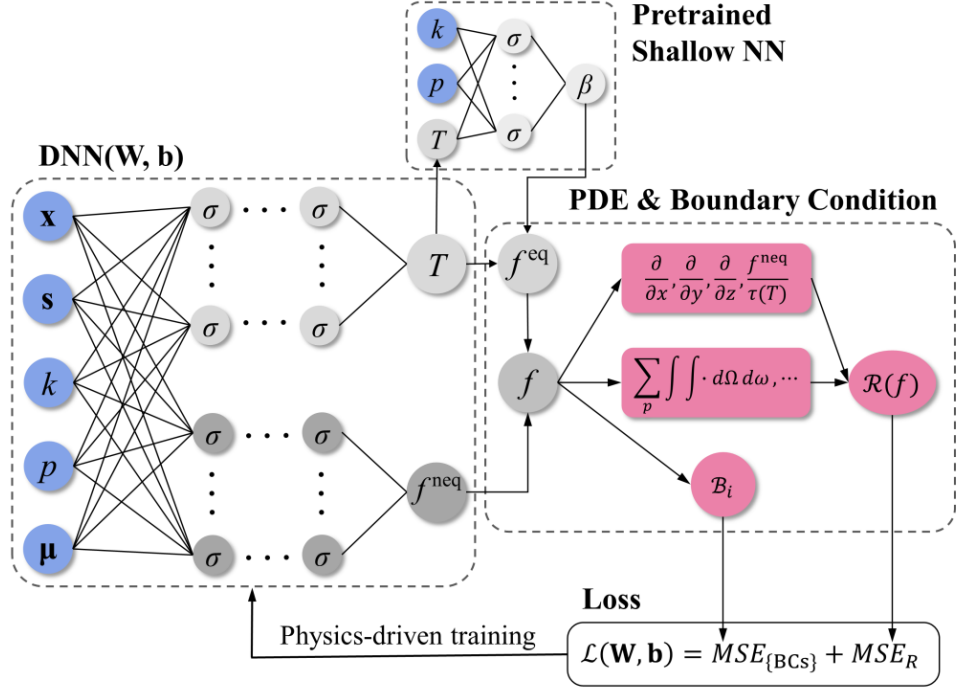

**Fig. 1: A schematic of the PINN framework for solving stationary phonon BTE with arbitrary temperature differences.** Two DNNs are employed to approximate the temperature ($T$) and non-equilibrium ($f^{\text{neq}}$) parts of the phonon distribution function, respectively. Inputs include spatial vector $x$, directional unit vector $\mathbf{s} = (cos\theta, sin\theta cos\varphi, sin\theta sin\varphi)$ ($\theta$ is the polar angle and $\varphi$ is the azimuthal angle), wave number $k$ and polarization $p$. $\boldsymbol{\mu}$ represents additional parameters, which is either characteristic length $L$ or boundary temperature difference $\Delta T$ in this study. $\sigma$ represents the activation function, which is set to be the Swish activation function. The pretrained shallow NN provides a scaling factor $\beta$ for approximating the equilibrium phonon distribution $f^{\text{eq}}$. The loss function contains residuals of the PDEs and boundary conditions on sampled collocation points in the simulation domain. The parameters in the two DNNs are learned by minimizing the total loss.

*Model Systems*

Numerical tests are carried out to evaluate the performance of this PINN scheme. We investigate several steady-state thermal transport problems with arbitrary temperature differences at different length scales, including 1D cross-plane, 2D in-plane, 2D rectangle and 3D cuboid. Single crystalline silicon is used as a model material, but the proposed model is applicable to other materials. We assume that the phonon dispersion relation of silicon is isotropic and the (100) direction information[9] is used in all tests. The derivations of phonon frequencies and relaxation times are based on Refs 10,48 (see the Methods section). Only one longitudinal acoustic (LA, set as $p = 1$) and two degenerate transverse acoustic (TA, set as $p = 0$) phonon branches are considered because the optical branches contribute little to the thermal transport[31]. It should be noted that including optical branches is possible and only involves expanding the sample space of the discrete input variable $p$. Since we assume that the silicon system is not heavily doped, electron-phonon interaction is not considered important in this work, but its effect can be easily included by adding its influence in the phonon relaxation times[49]. For each phonon branch, we discretize the wave vector space $k \in [0, 2\pi/a]$ equally into $N_k$ frequency bands by the midpoint rule, where $a = 5.431$ Å is the lattice constant for silicon. We set $N_k = 10$ in all cases as it gives a bulk thermal conductivity around 145.6 W m$^{-1}$ K$^{-1}$ at 300 K, which is in agreement with the literature value[50]. To obtain the average Knudsen number $\overline{\text{Kn}} = \bar{\lambda}/L$ at different temperatures, the average mean free path $\bar{\lambda}$ is introduced as

$$\bar{\lambda}(T) = \left(\sum_p \int_0^{\omega_{\max,p}} \hbar\omega D \frac{\partial f^{\text{eq}}}{\partial T} |\mathbf{v}|\tau d\omega\right)\left(\sum_p \int_0^{\omega_{\max,p}} \hbar\omega D \frac{\partial f^{\text{eq}}}{\partial T} d\omega\right)^{-1}. \quad (12)$$

The average mean free path as a function of temperature is shown in Fig. 2(a).

The training and testing details about numerical experiments are summarized in Table 1. $N_x$ is the number of interior points in the spatial domain (quasi-random Sobol sequences in training and uniform grids in testing), and $N_s$ is the number of solid angles by the Gauss-Legendre quadrature. $N_\mu$ represents the number of parameter values (length scale $L$ or boundary temperature difference $\Delta T$) sampled in a range given in Table 1. Then the total number of collocation points is $N_x \times N_s \times N_k \times N_p \times N_\mu$, where $N_p = 2$ is the number of phonon branches. The computation times and losses of training and testing processes are shown in Table 2, where the training loss is defined in Eq.

(10) after nondimensionalization, and the validation loss is the total loss evaluated in testing with the settings shown in Table 1.

**Table 1** Training and testing information of the numerical experiments. Two types of parametric training are used for 1D cross-plane problems with length scale $L$ and boundary temperature difference $\Delta T$, as denoted with superscript a and b.

| Case | $\mu$ | Range of $\mu$ | Training | | | Testing | | |
|---|---|---|---|---|---|---|---|---|
| | | | $N_x$ | $N_s$ | $N_\mu$ | $N_x$ | $N_s$ | $N_\mu$ |
| 1D cross-plane[a] | $L$ | $[10^{-8}, 10^{-4}]$ (m) | 40 | 16 | 9 | 80 | 64 | 17 |
| 1D cross-plane[b] | $\Delta T$ | $[200, 400]$ (K) | 40 | 16 | 5 | 80 | 64 | 5 |
| 2D in-plane | $L$ | $[10^{-8}, 10^{-4}]$ (m) | 400 | 100 | 5 | 1600 | 576 | 9 |
| 2D rectangle | $L$ | $[10^{-7}, 10^{-4}]$ (m) | 450 | 100 | 4 | 2601 | 576 | 4 |
| 3D cuboid | $L$ | $[3\times10^{-7}, 3\times10^{-6}]$ (m) | 1600 | 100 | 3 | 132651 | 576 | 4 |

**Table 2** Computation times and losses of numerical experiments. Wall times are recorded for evaluations of both temperature profiles and heat flux on a NVIDIA Tesla P100 GPU.

| Case | Training | | Testing | |
|---|---|---|---|---|
| | Loss | Wall time (h) | Loss | Wall time (s) |
| 1D cross-plane[a] | $2.0 \times 10^{-4}$ | 0.53 | $1.3 \times 10^{-3}$ | 0.11 |
| 1D cross-plane[b] | $1.0 \times 10^{-4}$ | 0.38 | $6.5 \times 10^{-4}$ | 0.05 |
| 2D in-plane | $4.5 \times 10^{-4}$ | 3.70 | $6.4 \times 10^{-4}$ | 7.70 |
| 2D rectangle | $8.5 \times 10^{-3}$ | 3.46 | $9.7 \times 10^{-3}$ | 5.79 |
| 3D cuboid | $8.0 \times 10^{-3}$ | 11.13 | $1.0 \times 10^{-2}$ | 301.26 |

*1D cross-plane phonon transport*

We first evaluate the quasi-1D cross-plane thermal transport in silicon films (Fig. 2(b)). The phenomena are described by an 1D phonon BTE with two isothermal boundary conditions (see the Methods section). The thickness of the film is $L$, and a temperature difference $\Delta T$ is induced in the $x$ direction. The temperature of the left boundary is set as $T_\text{L} = T_\text{ref} + \Delta T/2$, while that of the right boundary is set as $T_R = T_{ref} - \Delta T/2$. For the training of our PINN model, the spatial domain is equally discretized with 40 training points, and 16-point Gauss-Legendre quadrature is used for the phonon transport direction $\mathbf{s}_x = cos\theta$ (Table 1). Here we have conducted several tests at different reference temperatures $T_{ref}$, and the effects of $L$ and $\Delta T$ are separately studied by parametric learning.

To validate the proposed scheme, we perform numerical tests under the small temperature difference limit but without explicitly linearizing the equilibrium deviational distribution with respect to temperature. This is the limit where analytical solutions exist for such quasi-1D problems. Here, $T_{ref}$ is set to be 300 K and $\Delta T$ is set to 2 K. In this case, the pretrained scaling factor $\beta$ is almost equivalent to $\frac{\partial f^{eq}}{\partial T}$ since $\Delta T$ is sufficiently small. Parametric training is employed with thickness $L$ as an input to the DNNs. The model is trained with 9 samples of $L$ in the range between 10 nm and 100 μm. After training, the temperature and heat flux can be evaluated at any new $L$ given the interpolation ability of DNNs. Figure 2(c) shows the dimensionless temperature profiles $T^* = (T - T_R)/(T_L - T_R)$ at different $L$. The analytical solutions by the method of degenerate kernels[12] are included for comparison. Temperature profiles predicted by the PINN model are found to agree almost exactly with analytical solutions, with the discrepancy less than 1%. We also calculate the dimensionless thermal conductivities ($k_{eff}/k_{bulk}$), where $k_{eff} = qL/\Delta T$ is the effective thermal conductivity defined by Fourier's law, and $k_{bulk} = \frac{1}{3}\sum_p \int_0^{\omega_{max,p}} C|\mathbf{v}|^2 \tau d\omega$ is the bulk thermal conductivity in the diffusive limit. As shown in Fig. 2(d), we again observe good agreement for all testing points with the analytical solution (error < 0.7%). Although only trained with discrete thickness values, our model provides accurate predictions of thermal conductivity at unseen input thicknesses.

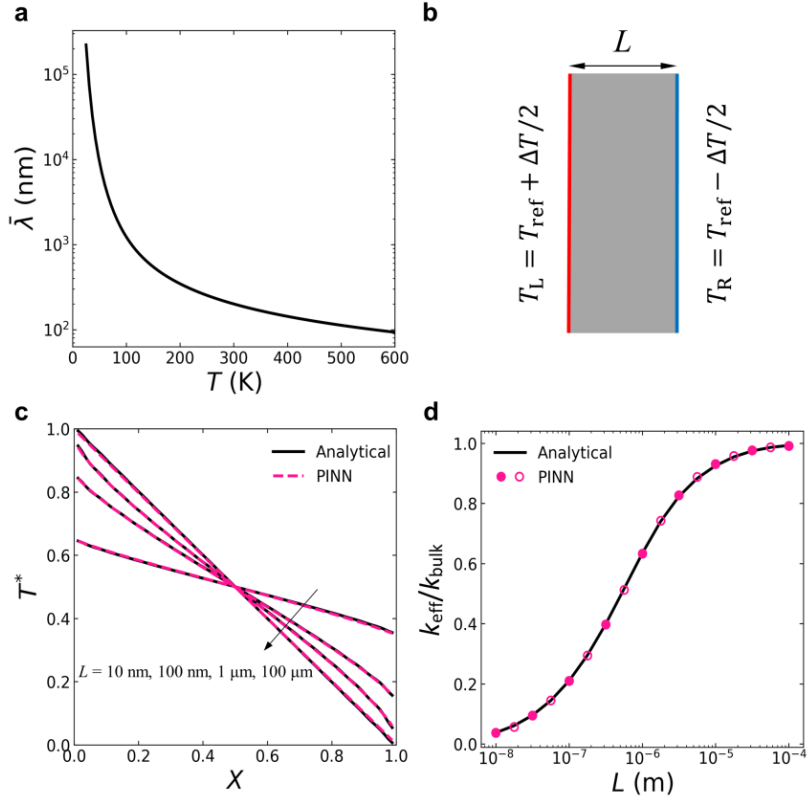

**Fig. 2: Results of 1D cross-plane phonon transport with small temperature differences.** (a) Temperature-dependent average phonon mean free path. (b) A schematic of the quasi-1D cross-plane phonon transport and the boundary temperatures. (c) Dimensionless temperature profiles of silicon thin films with different thicknesses ($L$ = 10 nm, 100 nm, 1 μm, 100 μm), where $T^* = (T - T_R)/(T_L - T_R)$ and $X = x/L$. The black solid lines represent analytical solutions to the quasi-1D phonon BTE. (d) Effective thermal conductivity normalized by the bulk thermal conductivity as a function of the thickness $L$. The filled circles represent the parameter points used in training, while the open circles are predicted points not included in training.

We then apply this framework to problems with larger temperature differences. In the ballistic limit, the phonon transport is governed by Stefan-Boltzmann law[51], where the phonon scattering is rare and the temperature across the whole system approximately follows $T^4 = (T_L^4 - T_R^4)/2$. We consider a silicon film with $T_L$ = 50 K, $T_R$ = 40 K and $L$ = 10 nm such that the average Knudsen number $\overline{Kn} = \bar{\lambda}/L \gg 10$. The shallow NN for $\beta$ has been trained for the relevant temperature range. We note that training this shallow NN is very fast, taking less than 20 seconds. Figure 3(a) shows the predicted temperature profile, which is very consistent with the analytical solution by Stefan-Boltzmann law. The small difference stems from the fact that there are still a small number of phonons with mean free path smaller than the thickness $L = 10$ nm, so they are close to diffusive.

As the system size increases, the phonon transport becomes more diffusive. Here, we continue to study the thermal transport in the diffusive regime at $T_{\text{ref}} = 300$ K, but with a much larger $\Delta T$. For fast predictions under different $\Delta T$, this time we incorporate $\Delta T$ as an input variable and learn the solutions in a parametric setting. Similar to the previous case, a shallow NN is trained to provide $\beta$ in the temperature range between 200 K and 400 K. Then, the PINN model is trained with 5 sampled $\Delta T$ values (20 K, 60 K, 100 K, 150 K and 200 K) at a fixed thickness $L = 100$ μm, where all phonons are expected to be diffusive. Figure 3(b) shows the predicted temperature profiles with different $\Delta T$. Compared to the analytical solutions based on Fourier's law, which is valid in the diffusive limit, our model accurately reproduces the analytical solutions (mean absolute error < 0.4 K) and captures the nonlinear effect due to the temperature-dependent thermal conductivity. Specifically, different from the linear profile in the diffusive regime under small $\Delta T$ (Fig. 2(c)), the temperature profile is predicted to be convex with larger $\Delta T$. Since at higher temperatures lattice thermal conductivity decreases due to the stronger intrinsic phonon scattering, the local temperature gradient increases with increasing temperature given the same heat flux. Thus, the convex temperature profile correctly indicates that the thermal conductivity decreases with the increasing temperature between 200 K and 400 K.

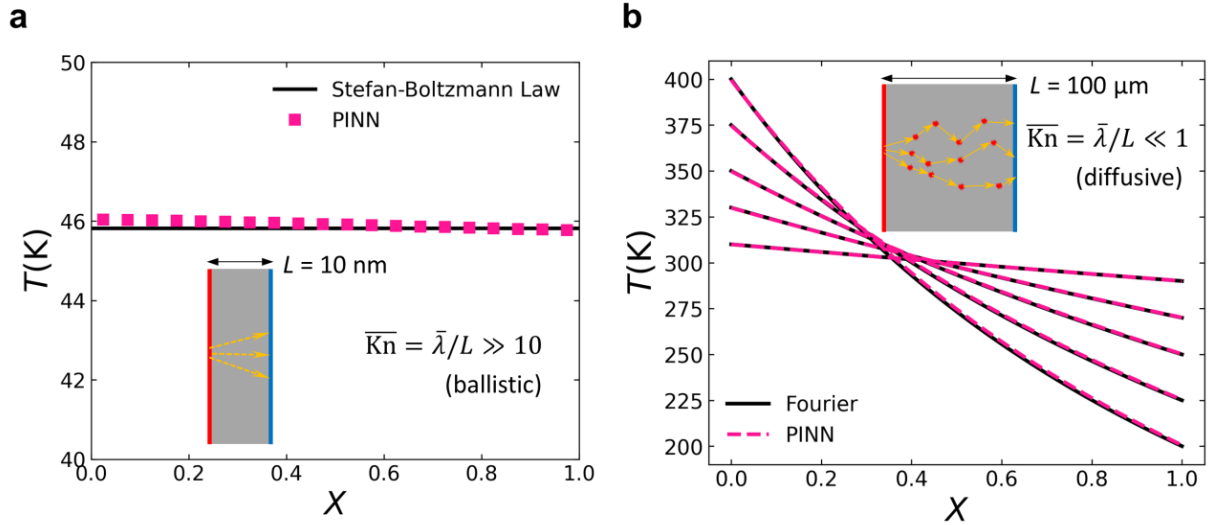

**Fig. 3: Results of 1D cross-plane phonon transport in the ballistic and diffusive limits with arbitrary temperature differences.** (a) Temperature profile of a silicon film at $L = 10$ nm, with boundary condition $T_\text{L} = 50$ K, $T_\text{R} = 40$ K, and $X = x/L$. The black solid line is the analytical solution by Stefan-Boltzmann law in the ballistic limit. (b) Temperature profiles with various boundary temperature differences at $L = 100$ μm, and the black lines are derived based on Fourier's law in the diffusive limit.

The same training procedure is adopted for cases at $T_\text{ref} = 100$ K, where the average mean free path changes more drastically (Fig. 2(a)). Considering the larger $\bar{\lambda}$ at lower temperatures, we set $L = 5$ mm to ensure that the phonon transport is close to the diffusive limit. As shown in Fig. 4(a), good agreement with Fourier solution in predicted temperature is confirmed for $\Delta T$ ranging from 10 K to 50 K. We also successfully reproduce the nonlinear temperature curves, which is infeasible under the assumption of small $\Delta T$. To further investigate the effects of $\Delta T$, we consider the cases near the ballistic regime with $L = 100$ nm. $T_\text{ref}$ is fixed at 100 K, while $\Delta T$ varies through a ratio $R = \Delta T/T_\text{ref}$ and is added as an input in parametric training. The dimensionless temperature profiles $T^*$ with different $R$ values are plotted in Fig. 4(b). Since the phonon transport in this case is dominated by the phonon boundary scattering, we observe temperature slips near the boundaries. As $\Delta T$ increases, the temperatures at the two boundaries increase, which can be attributed to the higher $\overline{\text{Kn}}$ at a lower temperature. The phonon boundary scattering has a larger impact at the cold boundary, leading to a larger temperature deviation from the predefined boundary temperature, and vice versa for the hot boundary.

All these testing cases confirm that the present scheme can not only describe the phonon transport correctly under small temperature ranges, but also provide accurate predictions when the temperature difference is large. As for the computational cost, the training time is estimated to be consistently less than one hour on a GPU in the form of parametric training, while all testing processes take less than one second (Table 2). It is important to note that since parameters like $L$ and $\Delta T$ can be added to the parametric space ($\boldsymbol{\mu}$) when training the model, a single training will allow the use of the model for different conditions (see Fig. 2(c-d), Fig. 3(b) and Fig. 4) – a main advantage of PINN over traditional numerical solvers, which need new simulations from scratch when any of the parameters are changed.

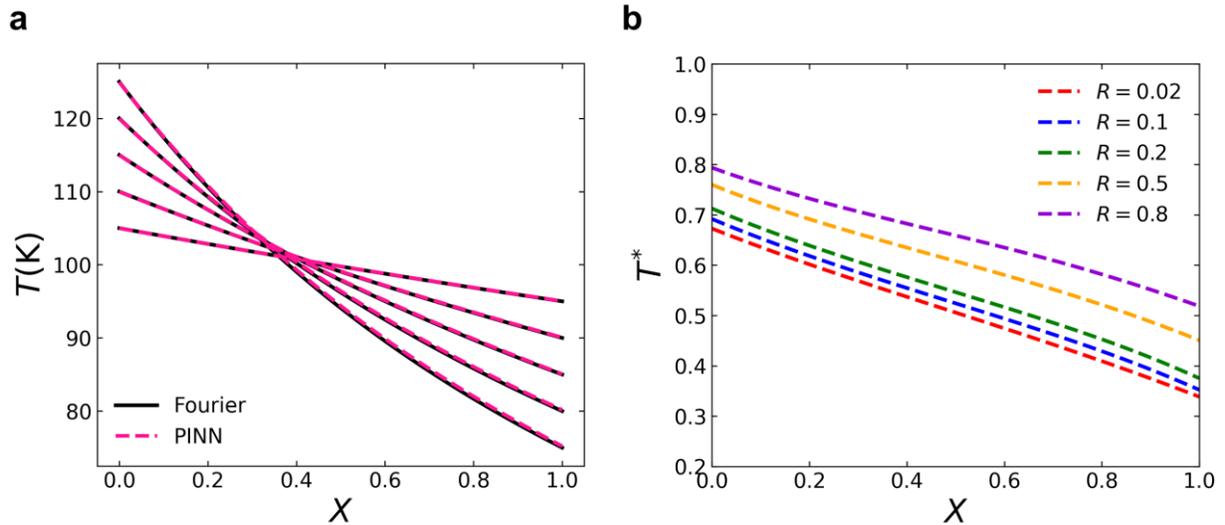

**Fig. 4: Results of 1D cross-plane phonon transport at $T_{\text{ref}} = 100$ K.** (a) Temperature profiles with different boundary temperature differences at $L = 5$ mm, and the black lines are derived based on Fourier's law. (b) Dimensionless temperature profiles with different boundary temperature differences at $L = 100$ nm, where $T^* = (T - T_{\text{R}})/\Delta T$, $R = \Delta T/T_{\text{ref}}$, $T_{\text{ref}} = 100$ K, and $X = x/L$.

*2D in-plane phonon transport*

For 2D in-plane thermal transport, we focus on the square silicon film (inset in Fig. 5) with a small temperature gradient ($\Delta T = 2$ K) along the x-direction, where analytical solutions can be derived from the Fuchs-Sondheimer theory for comparison[18]. Although approaching the small temperature gradient limit, we do not explicitly linearize the BTE but use the shallow NN for the scaling factor $\beta$. Diffusely reflecting boundary condition (see the Methods section) is applied to the top and bottom walls, and the other boundaries are periodic boundaries. The settings of the computational domain can be found in Table 1 and the Methods section. Here, $T_{\text{ref}}$ is fixed at 300 K, and the length scale $L$ is used as an input parameter. Two PINN models are trained to predict the phonon transport at $L$ within the range [10 nm, 1 μm] and [1 μm, 100 μm], respectively, to minimize the training loss for each range as phonon transport transitions from highly ballistic to diffusive.

Figure 5(a) shows the dimensionless x-directional heat flux $q_x^* = q_x(Y)/q_{\text{bulk}}$ at different $L$, where $q_{\text{bulk}} = -k_{\text{bulk}} \cdot \Delta T/L$. The differences between the PINN predictions and the analytical solutions are less than 2.9%. We also observe a good agreement (error < 1.9%) in effective thermal conductivity $k_{\text{eff}} = -(dT/dx)^{-1} \int_0^1 q_x(Y)dY$ as shown in Fig. 5(b), and again the present method reproduces the varying effective thermal conductivity due to the change of length scale. Like our previous model devised for small temperature differences[46], the present scheme shows high accuracy in 2D in-plane thermal transport. The evaluation is also very fast and takes less than 8 seconds on a domain with much more collocation points than the training domain (see Tables 1 and 2).

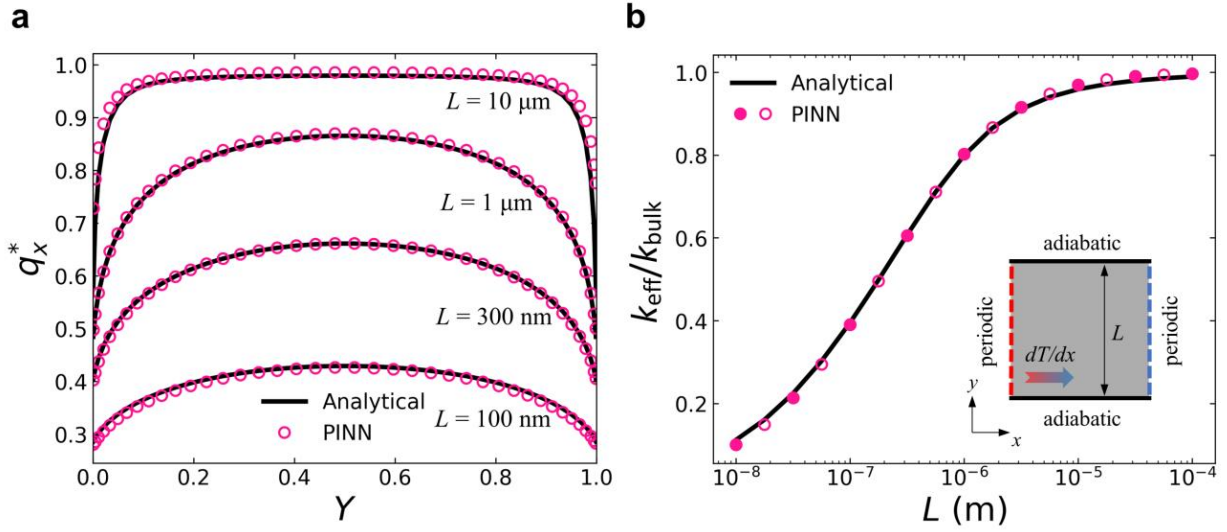

**Fig. 5: Results of 2D in-plane phonon transport ($\Delta T = 2$ K).** (a) Dimensionless x-directional heat flux results along the y-axis (see inset in panel b) in silicon films with different length scales, where $q_x^* = q_x(Y)/q_{\text{bulk}}$ and $Y = y/L$. From bottom to top, the PINN predictions are shown as circles for $L$ = 100 nm, 300 nm, 1 μm, 10 μm. The black solid lines represent analytical solutions by the Fuchs-Sondheimer theory. (c) Effective thermal conductivity normalized by the bulk thermal conductivity at different length scales. The filled circles represent the parameter points used in training, while the open circles are predicted points not included in training. The inset shows the schematic of the simulation domain and boundary conditions.

*2D rectangle phonon transport*

Next, we apply our method to the phonon transport in a 2D rectangle domain (Fig. 6(a)) with large temperature differences. The length and width of the geometry are $L$ and 0.5$L$, respectively. To mimic the condition of Joule self-heating, we apply a Gaussian temperature distribution $T_h$ to the top boundary, while other boundaries are held at a lower temperature $T_c = 300$ K. The Gaussian temperature distribution is set with the full width at half maximum (FWHM) to be 0.4$L$, and the difference between the peak temperature at the center of the top wall $T_{\text{max}}$ and $T_c$ is 100 K.

Parametric training is conducted on 4 geometries with $L$ ranging from 100 nm to 100 μm. It is observed that PINN successfully reproduces the Gaussian temperature distribution at the top boundary (Fig. 6(b)). Figures 6(c-e) show the predicted 2D temperature contours at different $L$.

Although there is no available analytical solution for direct comparison, we derive a Fourier solution (i.e., solution in the diffusive limit) using a simple PINN model for a 2D steady-state heat equation (Fig. 6(f)). This result can be treated as the benchmark solution in the diffusive limit as the final training loss is as low as $4\times10^{-4}$.

We can find that the result of the 100 μm case (Fig. 6(e)), which is close to the diffusive limit, is nearly identical to the Fourier benchmark (Fig. 6(f)), with the mean absolute error less than 0.3 K. Obvious temperature slips near the top boundary are also observed in cases with smaller $L$, and as $L$ increases the slip decreases. This test confirms the capability of the present scheme in solving 2D phonon transport under large temperature gradients. Another feature about this scheme is that although the training time increases due to more training points used and the higher input dimension (Table 2), the evaluation cost can be less than one second if we only need to predict the temperature profile, which is the output of a sub-network in our model (see Fig. 1).

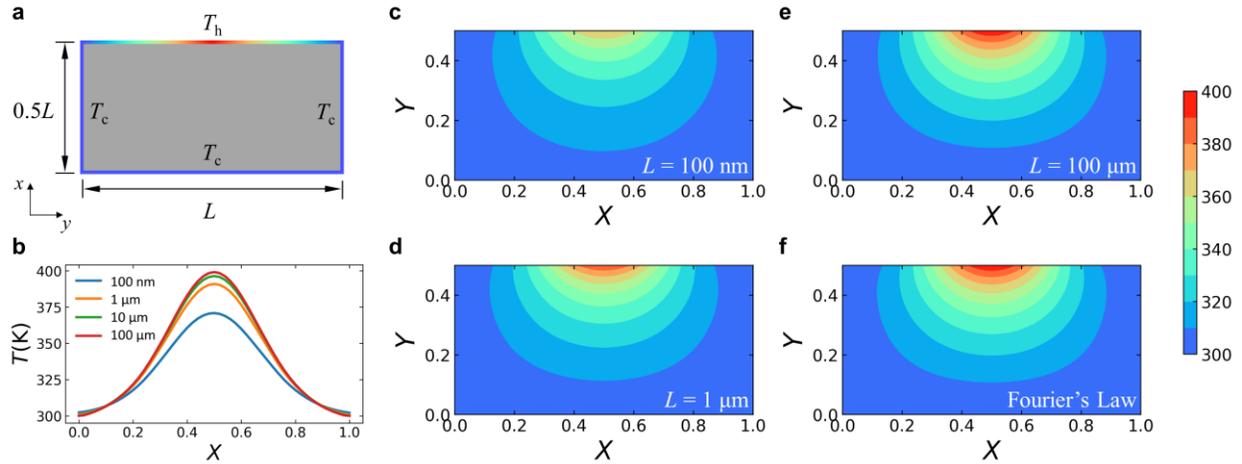

**Fig. 6: Results of 2D rectangle phonon transport.** (a) The computational domain is of size $L \times 0.5L$. Gaussian temperature distribution $T_h$ is applied to the top boundary (with $T_{max} = 400\ K$), while all the other boundaries are maintained at a lower temperature ($T_c = 300\ K$). (b) PINN-predicted temperature distributions at the top wall at different length scales. (c-e) Predicted temperature contours at length scale $L = 100$ nm, 1 μm, 100 μm. (f) Solution of the 2D heat equation based on the Fourier's law, which is obtained by a well-trained PINN. $X$ and $Y$ are normalized spatial coordinates.

*3D cuboid phonon transport*

To validate the present model for problems in a more realistic setting, we consider the phonon transport in a 3D cuboid geometry as an extension of the last 2D case. The test geometry is a silicon block of size $L \times L \times 0.5L$, with a circular hot spot following a Gaussian temperature distribution (FWHM = $0.4L$) on the top surface, as depicted in Fig. 7(a). To demonstrate that the present model is applicable to 3D problems where the phonon mean free path features a wide span of orders of magnitudes (e.g., below 200 K), we select a different temperature range than that in the 2D case in this test. The peak temperature $T_{\max}$ is set to 200 K at the center of the top surface, and the other surfaces are maintained at $T_c = 100$ K. A PINN model is first trained at $L = 1$ mm without additional input parameters. Similar to the previous test, we compare the results to the Fourier solution for the 3D heat equation from a PINN model (training loss $< 7\times 10^{-4}$).

Figure 7(b) shows the temperature contours in two central planes predicted by our PINN model without parametric learning, which are evaluated on a computational domain with much more points than the training domain. Comparing our prediction with the Fourier benchmark in the plane at $y = 0.5L$ (Fig. 7(c-d)), we find that the difference is small as the 1 mm case is close to the diffusive limit, while the mean absolute temperature difference is less than 0.2 K across the whole system. We have also performed a parametric training with variable $L$ sampled in the range between 300 nm and 3 μm (Table 1) and achieved the ability to predict the temperature profile for various sample lengths within a few minutes. Figure 7(e-f) show the predicted temperature profiles at two length scales, while $L = 500$ nm is a predicted point not included in the training. Based on the good performance in 1D and 2D problems, we expect high computational accuracies for 3D geometries of different sizes as well, but there are no benchmark results to compare with for the 3D non-diffusive cases. It is noted that such fast and accurate prediction in 3D geometries has not been achieved by any other methods under large temperature gradients.

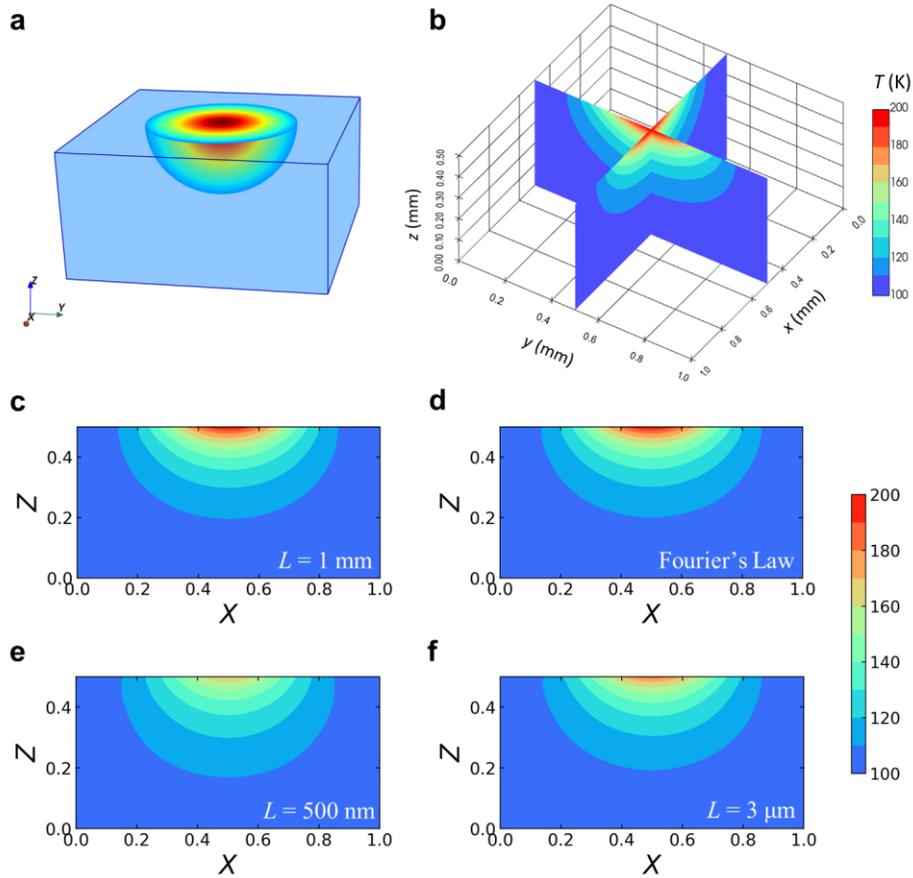

**Fig. 7: Results of 3D cuboid phonon transport.** (a) Schematic of the 3D thermal transport in a cuboid geometry of size $L \times L \times 0.5L$. Gaussian temperature distribution is applied to the top surface ($T_{\max} = 200$ K), while all the other surfaces are maintained at a lower temperature ($T_c = 100$ K). (b) Predicted steady-state temperature contour for a 3D system of size 1 mm × 1mm × 0.5 mm. (c) Predicted temperature contour in the central plane ($y = 0.5L$) at $L = 1$ mm. (d) Solution of 3D heat equation based on Fourier's law under the same boundary conditions, which is obtained by a well-trained PINN. (e-f) Predicted temperature contours at two length scales through parametric training, and $L = 500$ nm is not included in the training. $X$ and $Z$ are normalized spatial coordinates.

**Discussion**

In summary, a deep learning-based PINN model is developed for solving mode-resolved phonon BTE with arbitrary temperature differences. Numerical tests show that the present scheme can accurately predict steady-state phonon transport from 1D to 3D under arbitrary temperature differences, which is computationally challenging, if not impossible, for conventional numerical methods. Under large temperature gradients, the phonon transport is found to be very different from that under small temperature differences due to the temperature-dependent phonon relaxation times. When the temperature difference is large, the phonon relaxation time can vary significantly over space for a given frequency and polarization depending on the local temperature, and the present scheme successfully handles such situations through the introduction of a scaling factor described by a pretrained shallow NN. Parametric learning is also enabled by including the length scale or boundary temperature difference as additional inputs to the model, allowing for efficient investigation of effects due to varying temperature difference and Knudsen numbers. As for the computational cost, we note that conventional solvers usually require long computation time (tens of hours) and large memory (hundreds of gigabytes) for 3D solutions under large temperature difference even using large scale parallel computing[29,52]. However, the present model can provide solutions at any point in the computational domain within at most several minutes, and it is very computationally efficient even considering the training time due to the implementation of parametric learning.

The parametric learning feature, together with the low evaluation cost, may allow for efficient search in parameterized spaces for design purposes. To apply this scheme to the simulations of realistic device-level thermal transport, we need the phonon dispersion relations for the materials constituting the target system. When the component materials are heavily doped, the effects of electron-phonon interaction should be included by adjusting the phonon relaxation time accordingly. Besides, an efficient sampling strategy must be adopted for the improved learning performance when the system structure is more complex. Since the current scheme approximates the solution function by minimizing the BTE residuals on the sampled collocation points, it is theoretically applicable to inhomogeneous systems such as heterojunctions and porous materials given the intrinsic phonon properties and interfacial phonon scattering. This PINN scheme is also easy to implement and does not need any labeled data for training. It can be a powerful tool in studying multiscale thermal transport for applications like thermoelectrics and electronics thermal management.

While being accurate and efficient in predicting multiscale thermal transport, the current scheme still has limitations, which warrants further research. In particular, our framework is designed for steady-state problems, and modifications are required in order to capture the transient thermal transport. For example, Long-Short Term Memory (LSTM) recurrent neural network architecture[53] could potentially be used to deal with dynamic systems. Furthermore, for realistic electronic device-level simulations, it is desirable to solve the phonon BTE and electron BTE simultaneously. Since most existing methods for electro-thermal simulations have employed simplifications in physics equations[54] or separate solvers for electrons and phonons[55,56], a unified PINN model would be ideal for reliable investigation of self-heating effects by solving the coupled BTEs.

## Methods

*PINN architecture and training*

The proposed PINN model consists of two DNNs for training and one pretrained ANN, where two DNNs have the local temperature (*T*) and the non-equilibrium ($f^{\text{neq}}$) part of the phonon distribution function as output, respectively (Fig. 1). With 30 neurons per layer, the DNN for $f^{\text{neq}}$ has a structure of 8 hidden layers, while the DNN for *T* has a varying number of hidden layers depending on the problem dimension. The pretrained ANN for scaling factor *β* has only one hidden layer with 30 neurons. Two DNNs are trained simultaneously with a unified physics-informed loss function. We employ the Swish activation function (*x*·Sigmoid(*x*))[57] in each layer except the last one, where a linear activation function is applied. The Adam optimizer[58], a robust variant of the stochastic gradient descent algorithm, is used to solve the optimization problem defined in Eq. (10) by training on mini-batches of inputs. The initial learning rate is set as $5 \times 10^{-3}$, and training points are generated by sampling of the input domain. To approximate the integrals in Eq. (8), Gauss-Legendre quadrature[59] is adopted for the solid angle space, while the midpoint rule is used for the frequency space. In the case the spatial domain is logically rectangular, we can set the interior training points as quasi-random low-discrepancy Sobol sequences[60] to alleviate the curse of dimensionality. Input spatial coordinates are scaled to the range [0, 1]. The PINN algorithm is implemented within the PyTorch platform[61], and all numerical experiments are performed on a single NVIDIA Tesla P100 Graphic Processing Unit (GPU).

*Boundary conditions*

Three categories of boundary conditions are usually met in phonon transport problems, including isothermal boundary condition, diffusely reflecting boundary condition and periodic boundary condition. These boundary conditions can be applied to problems with any parameter sets **μ**.

*Isothermal boundary* absorbs all incident phonons and emits phonons in thermal equilibrium with the boundary temperature $T_b$. Mathematically, this can be expressed as

$$f(\mathbf{x}_b, \mathbf{s}, k, p) = f^{eq}(k, p, T_b), \qquad \mathbf{s} \cdot \mathbf{n}_b > 0, \tag{13}$$

where $\mathbf{n}_b$ is the normal unit vector pointing into the simulation domain.

*Diffusely reflecting boundary* is a type of adiabatic boundary. At this boundary, the net heat flux is zero, meaning that the phonons are reflected with equal probability along all possible directions, namely,

$$f(\mathbf{x}_b, \mathbf{s}, k, p) = \frac{1}{\pi} \int_{\mathbf{s}' \cdot \mathbf{n}_b < 0} f(\mathbf{x}_b, \mathbf{s}', k, p) |\mathbf{s}' \cdot \mathbf{n}_b| d\Omega, \qquad \mathbf{s} \cdot \mathbf{n}_b > 0. \tag{14}$$

For the *periodic boundary*, a phonon that crosses it is emitted at the opposite boundary with the same velocity vector and frequency. Besides, two corresponding boundaries follow the local thermal equilibrium,

$$f(\mathbf{x}_{b_1}, \mathbf{s}, k, p) - f^{eq}(k, p, T_{b_1}) = f(\mathbf{x}_{b_2}, \mathbf{s}, k, p) - f^{eq}(k, p, T_{b_2}), \tag{15}$$

where $\mathbf{x}_{b_1}$, $T_{b_1}$ and $\mathbf{x}_{b_2}$, $T_{b_2}$ are the spatial coordinates and temperatures of two associated periodic boundaries $b_1$ and $b_2$, respectively.

*Phonon dispersion and scattering*

The dispersion relations of the acoustic phonons are approximated as $\omega = c_1 k + c_2 k^2$ [48], where for LA branch $c_1 = 9.01 \times 10^5$ cm/s, $c_2 = -2.0 \times 10^{-3}$ cm$^2$/s; for TA branch $c_1 = 5.23 \times 10^5$ cm/s, $c_2 = -2.26 \times 10^{-3}$ cm$^2$/s. The Matthiessen's rule is used to estimate the effective relaxation time by combining different scattering processes[62], including the impurity scattering, umklapp (U) and normal (N) phonon-phonon scattering, $\tau^{-1} = \tau_{\text{impurity}}^{-1} + \tau_{\text{U}}^{-1} + \tau_{\text{N}}^{-1} = \tau_{\text{impurity}}^{-1} + \tau_{\text{NU}}^{-1}$, where the relaxation time formulas and coefficients[10] are given in Table 3. We note that the dispersion and relaxation times can also be from first-principles calculations for each discrete mode[63]. The PINN implementation will not change.

**Table 3** Relaxation time formulas and coefficients.

| | |
|---|---|
| $\tau_{\text{impurity}}^{-1}$ | $A_i \omega^4$, $A_i = 1.498 \times 10^{-45}$ s$^3$ |
| LA | $\tau_{\text{NU}}^{-1} = B_{\text{L}} \omega^2 T^3$, $B_{\text{L}} = 1.180 \times 10^{-24}$ K$^{-3}$ |
| TA | $\tau_{\text{NU}}^{-1} = B_{\text{T}} \omega T^4$, $0 \leq k < \pi/a$ |
| | $\tau_{\text{NU}}^{-1} = B_{\text{U}} \omega^2 / \sinh(\hbar\omega/k_{\text{B}}T)$, $\pi/a \leq k \leq 2\pi/a$ |
| | $B_{\text{T}} = 8.708 \times 10^{-13}$ K$^{-3}$, $B_{\text{U}} = 2.890 \times 10^{-18}$ s |
| | $a = 0.5431$ nm |

## Data availability

All data needed to evaluate the conclusions in the paper are present in the paper. Additional data related to this paper may be requested from the authors.

## Code availability

The code required to reproduce these findings are available to download from https://github.com/RuiyangLi6/PINN-pBTE-deltaT upon publication.

## Acknowledgements

The authors would like to thank ONR MURI (N00014-18-1-2429) for the financial support. The simulations are supported by the Notre Dame Center for Research Computing, and NSF through the eXtreme Science and Engineering Discovery Environment (XSEDE) computing resources provided by Texas Advanced Computing Center (TACC) Stampede II under grant number TG-CTS100078. This work is also supported by the National Research Foundation of Korea (NRF) grant funded by the Korea government (MSIT) (No. NRF-2021R1C1C1006251).

## Author contributions

R.L., E.L. and T.L. conceived the idea and initiated this project. R.L. designed and trained the model. All authors have contributed to the writing of the paper and the analysis of the data.

## Competing interests

The authors declare no competing interests.